\documentclass[a4paper,10pt]{article}
\usepackage[left=2cm,bottom=2cm,right=2cm,top=1cm]{geometry}
\usepackage[T1]{fontenc}
\usepackage{graphicx}
\usepackage{booktabs}
\usepackage{verbatim}
\usepackage{amsmath,amsfonts}

\graphicspath{ {./images/}}

\newcommand\T{\mathcal T_{\mathbf k}}
\newcommand\TT{\mathcal T_{\mathbf k}^*}
\newcommand\Dt{\mathcal D_\theta}
\newcommand\Dk{\mathcal D_{\theta, \mathbf k}}
\newcommand\rec{\widehat  {\mathbf x}}
\newcommand\als{\widetilde{\mathbf x}}

\title{\textbf{LSST: Learned Single-Shot Trajectory and \\Reconstruction Network for MR Imaging}}

\author{Hemant Kumar Aggarwal$^1$ \and Sudhanya Chatterjee$^1$ \and Dattesh Shanbhag$^1$ \and Uday Patil$^1$ \and K.V.S.~Hari$^2$ }

\date{
	$^1$Advanced Technology Group, AI Organization, GE HealthCare, Bangalore India \\ \texttt{ \{hemantkumar.aggarwal,sudhanya.chatterjee,dattesh.shanbhag,uday.patil\} @gehealthcare.com}  \\%
	$^2$Center for Brain Research, Indian Institute of Science, Bangalore, India \\ \texttt{hari@iisc.ac.in}\\[2ex]%
	}

\begin{document}
\maketitle            
\begin{abstract}
Single-shot magnetic resonance (MR) imaging acquires the entire k-space data in a single shot and it has various applications in whole-body imaging. However, the long acquisition time for the entire k-space in single-shot fast spin echo~(SSFSE) MR imaging poses a challenge, as it introduces T2-blur in the acquired images. This study aims to enhance the reconstruction quality of SSFSE MR images by (a) optimizing the trajectory for measuring the k-space, (b) acquiring fewer samples to speed up the acquisition process, and (c) reducing the impact of T2-blur. The proposed method adheres to physics constraints due to maximum gradient strength and slew-rate available while optimizing the trajectory within an end-to-end learning framework. Experiments were conducted on publicly available fastMRI multichannel dataset with 8-fold and 16-fold acceleration factors. An experienced radiologist's  evaluation on a five-point Likert scale indicates improvements in the reconstruction quality as the ACL fibers are sharper than comparative methods.% and meniscus region is sharper everywhere.		

%\keywords{Trajectory Optimization \and Single-Shot MRI \and Deep Learning.}
\end{abstract}

\section{Introduction}
Magnetic Resonance~(MR) Imaging is a non-invasive technique that offers superior soft-tissue contrast compared to other imaging modalities such as X-ray or CT.  
The quality of reconstructed images are influenced not only by the number of samples acquired but also by the sampling scheme employed. Therefore, optimizing the sampling pattern for acquisition is critical to further enhance the reconstruction quality \cite{senel2019,loupe,haldar2019oedipus,Reeves2000,sherry2019,gozcu2018learning,samsonov,waveloraks,jmodl}.

Prior work exist regarding sampling pattern optimization scheme independent of reconstruction algorithm~\cite{senel2019,haldar2019oedipus,Reeves2000}, active sensing techniques~\cite{jin,Zhang} which optimizes sampling location after each TR, and methods that focus on multi-shot fast spin echo (FSE) sequences such as T2-Shuffling~\cite{t2shuffling}. This study focuses on jointly optimizing sampling pattern and reconstruction network for single-shot spin echo MR imaging sequences. 

The process of undersampling k-space in two dimensions to expedite MR acquisition cannot be arbitrary due to practical considerations such as system configuration of gradient and slew-rate. Random undersampling, which necessitates swift gradient switching, is often impractical as it results in high eddy current-related artifacts~\cite{csreview}. When formulating a k-space trajectory, it's crucial to consider the relevant MR system hardware constraints, specifically, the maximum gradient magnitude ($G_{\max}$) and maximum slew-rate ($S_{\max}$). These constraints limit the maximum velocity ($v_{\max} = \gamma G_{\max}$)  and acceleration ($a_{\max} = \gamma S_{\max}$) of the trajectory, where $\gamma$ is the Gyromagnetic ratio~\cite{buxton2009introduction}.

Given the benefits of arbitrary single-shot trajectories, this work focuses on optimizing them for a MR system with specific gradient and slew-rate specifications. This is achieved by determining the shortest path through a random initial set of points, akin to solving a Traveling Salesman Problem (TSP)~\cite{chauffert2013travelling}. A TSP trajectory might not meet the gradient constraints but provides a good initialization for the trajectory in the joint optimization framework.

Our approach improves upon the PILOT~\cite{pilot} study by eliminating the need for solving an explicit TSP during joint optimization. Unlike PILOT, this work explicitly accounts T2-Blur in single-shot acquisition and enhances reconstruction quality by incorporating the Density Compensation Factor~(DCF) during training, and initializing the network input with a model-based solution. Unlike the BJORK~\cite{bjork} study, which doesn't account for T2-blur, which is relevant for single-shot MR imaging, our method account for it. We propose a direct-inversion network to account for the unknown T2-Blur. Key contributions of this work include:
\begin{itemize}
	\item Proposing  an approach to accelerate 2D spin echo (SE) MR imaging as single-shot fast spin echo imaging, offering an effective alternative to the 1D undersampling approach that produces considerable aliasing and blurring artifacts for high acceleration rates.
	\item Development of a framework for accelerated single-shot MR image acquisition and reconstruction with a focus on reducing T2-Blur, noise, and aliasing artifacts occuring in single-shot fast spin echo imaging.        	
	\item Handling an unknown forward model in single-shot acquisition due to unknown T2-Blur, using a direct-inversion network. The T2 blur is simulated for the proposed imaging method in a physics aware manner.	
\end{itemize}

\section{Proposed Generalized Framework}
\begin{figure}
	\centering
	\includegraphics[width=0.95\textwidth]{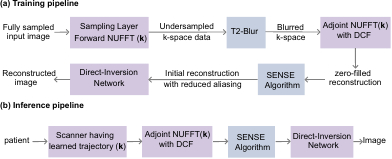}		
	\caption{The training~(a) and inference~(b) pipeline of the proposed joint optimization framework. Here purple blocks contain trainable parameters. }
	\label{fig:model}
\end{figure}

Figure~\ref{fig:model} illustrates the training and inference pipeline of our proposed Learned Single-Shot Trajectory (LSST) optimization framework, which concurrently optimizes both the k-space trajectory and the reconstruction network. Section~\ref{sec:acq} discusses trainable single-shot image acquisition model and section~\ref{sec:recon} discusses  proposed joint optimization framework for trajectory optimization.

\subsection{Trainable Single-shot Acquisition Model}
\label{sec:acq}

\begin{figure}
	\centering	
	\includegraphics[width=0.9\linewidth]{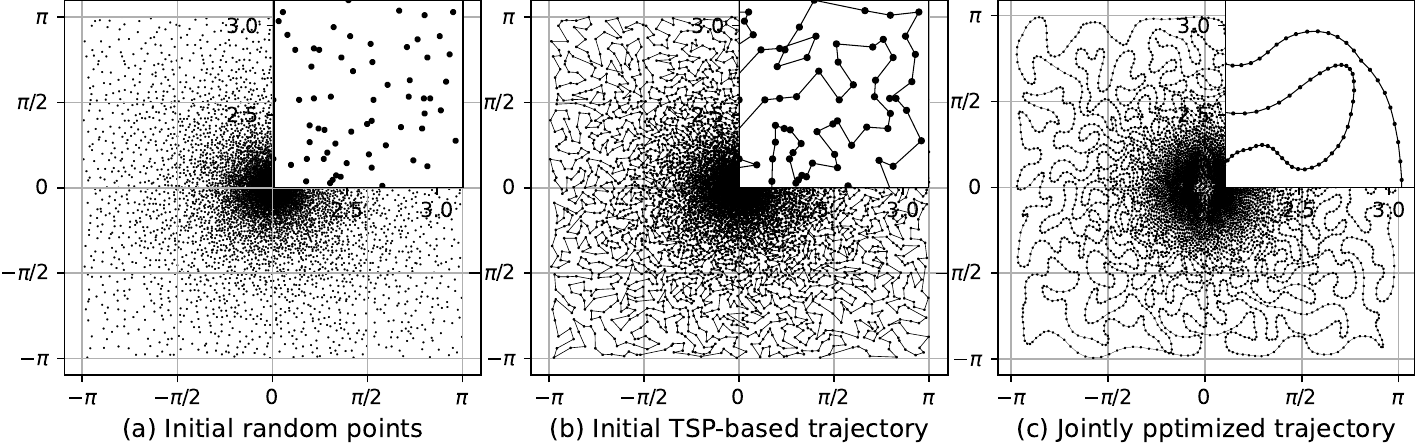}	
	\caption{This figure (a) shows an example of a random set of points acquired to achieve an acceleration factor R=8. (b) shows corresponding TSP-based trajectory that does not satisfy physics constraints on maximum gradient strength and slew-rate. (c) shows optimized trajectory using the proposed LSST framework that has smooth curvature as seen from zoomed portions in range [$2\pi/3 ,\pi$] that optimized trajectory has smooth curvature compared to (b) since it satisfy physics constraints.}
	\label{fig:traj}	
\end{figure}

Initially a random variable-density sampling mask for a given acceleration factor $R$ is generated, as depicted in Fig.~\ref{fig:traj}(a).  An initial trajectory~$\bf k$ is generated from these random points using a TSP solver, such that the k-space center is sampled as the  first point. The trajectory defines the order in which k-space is traversed to acquire the data in a single-shot. The network is initialized with this trajectory shown in Fig.~\ref{fig:traj}(b). A direct TSP-based trajectory is infeasible since it does not satisfy physics constraints based on system hardware. Additionally, this generic trajectory is not necessarily optimal for a given dataset/task. Therefore, optimization of the trajectory is based on a given dataset while enforcing the physics constraints. Thus, trajectory is treated as a trainable parameter~$\bf k$. After imposing these constraints, the learned trajectory has relatively smooth curvatures which are  evident from Fig.~\ref{fig:traj}(c). 

As shown in Fig.~\ref{fig:model}(a), we input a fully sampled image $\mathbf x$ to a backpropagation-compatible non-uniform Fourier transform~(NUFFT)~\cite{bjork} operator~$\T$ to generate the undersampled k-space measurements $\mathbf y$  using the forward model
\begin{equation}
	\label{eq:forward}
	\mathbf y=\T (\mathbf x) +\bf n,
\end{equation}
where $\bf n$ is an additive Gaussian noise. Here, the operator $\T$ consists of k-space sampling trajectory ($\mathbf k \in \mathbb{C}^m $), non-uniform Fourier transform, and coil sensitivity maps~(CSM).  This forward model is often considered in most image reconstruction and joint optimization frameworks including:~\cite{varnet,loupe,admmnet,bjork,pilot,varnet,dagan}.  However, this model~\eqref{eq:forward} does not account for the T2-blur that is present in the single-shot MR acquisition. We can model this unknown tissue specific T2-blur in k-space as a following modulation operation
\begin{equation}
	\label{eq:blur}
	\mathbf y_b=\mathcal B \odot	\mathbf y,
\end{equation}
where $\mathcal B$ represents the unknown k-space blur modulation function, and $\mathbf y_b$ represent modulated, undersampled, and noisy measurements. Combining,~ \eqref{eq:forward} and~\eqref{eq:blur}, we get relatively accurate forward model for single-shot acquisition as
\begin{equation}
	\label{eq:ssmodel}
	\mathbf y_b=\mathcal B \odot \left(\T (\mathbf x) \right) +\mathbf n.
\end{equation}
The next section describes the reconstruction process from these acquired blurred noisy undersampled measurements $\mathbf y_b$.

%--------------------------------------------------------------------------

\subsection{Proposed Image Reconstruction Pipeline}
\label{sec:recon}

This section describes the proposed three-step reconstruction process that involves initial iterative Sensitivity Encoding~(SENSE) based reconstruction followed by deep learning based reconstruction and artifact reduction step. 

\subsubsection{Initial SENSE Reconstruction}
Given that the trajectory densely samples the center of k-space and sparsely samples the periphery, it can introduce a bias towards the lower frequencies, resulting in a blurred reconstruction from the k-space measurements. Therefore,we explicitly account for density compensation prior to the reconstruction process.  This work utilizes a backpropagation compatible implementation of a density compensation algorithm~\cite{dcfPipe} on blurred measurements, $\mathbf y_b$, to obtain an approximation of regridding reconstruction, $\als$  using conjugate of the acquisition operator as  $\als=\TT(\mathbf y_b)$. 

The occurrence of blur in single-shot acquisition presents additional challenges due to the pronounced undersampling artifacts and get exacerbated at higher acceleration factors. Additionally, an unknown modulation function contributes to the blur in the reconstructed images (compounding the aliasing artifacts). The density compensated regridding reconstruction $\als$, obtained from blurred measurements $\mathbf y_b$, can be improved by using reconstruction algorithms such as  Sensitivity Encoding (SENSE) that solves the following $\ell_2$-regularized optimization problem 
\begin{equation}
	\label{eq:sense}
	\text{arg}\min_x \|\mathbf y_b -\T(\mathbf x) \|_2^2 + \lambda \|\mathbf x\|_2^2.
	%	\mathbf y=\T (\mathbf x) +\bf n.
\end{equation}
iteratively using Conjugate-Gradient algorithm, whose solution can be represented as
\begin{equation}
	\label{eq:sensesol}
	\mathbf x_0= (\T^*\T+\lambda \mathcal I)^{-1}(\als).	
\end{equation}
Although, due to the absence of blur information, single-shot forward model in~\eqref{eq:ssmodel} can not be directly used to obtain the solution $\mathbf x_0$, still it acts as a good initial guess to the neural network since the aliasing artifacts are significantly reduced in  $\mathbf x_0$ as compared to~$\als$. 

\subsubsection{Direct-Inversion Network}
Once we have an initial solution $\mathbf x_0$, we can reconstruct the final image using a CNN. Unlike BJORK~\cite{bjork}, the single-short forward model~\eqref{eq:ssmodel} has unknown blur therefore a model-based deep learning approach is not directly applicable for the single-shot reconstruction and trajectory optimization. Therefore, in this work, we propose to reduce the artifacts present in~$\mathbf x_0$ using a direct-inversion network such UNet~\cite{unet}. It is possible to utilize a fixed feasible trajectory such as~\cite{cap} and obtain the reconstructed image $\rec$ from~$\mathbf x_0$ using  deep learning methods as
\begin{equation}\label{eq:capdl}
	\rec=\Dt(\mathbf x_0)
\end{equation}
Here, $\Dt$ is a CNN with trainable parameters $\theta$. This method only optimizes the network parameters once the samples are acquired. However, the joint optimization of the sampling trajectory ($\mathbf k$) and reconstruction network parameters ($\theta$)  can further improve the reconstruction quality. The joint optimization can be represented as
\begin{equation}\label{eq:optdl}
	\rec=\Dk(\als),
\end{equation}
where, $\Dk$ represent the network that jointly optimizes trajectory and reconstruction parameters using the following loss function
\begin{equation}
	\label{eq:simpleloss}
	\{\theta^*,\mathbf k^*\}=\text{arg}\min_{\theta,\mathbf{k}}	L_{\text{total}} (\rec,\mathbf x).
\end{equation}
Here, $\mathbf x$ represents the ground truth data and $L_{\text{total}}=L_{\text{task}} + \beta L_{\text{const}}$ is the total loss consisting of task loss~$L_{\text{task}}$ and constraint loss~$L_{\text{const}}$ with $\beta$ as hyperparameter.
$L_{\text{task}}$ is a task-based loss function to minimize the error between ground truth and the reconstructed images. The $L_{\text{task}}$  was considered as
\begin{equation}
	\label{eq:ltask}
	L_{\text{task}} = \sum_{i=1}^{N}  \left ( || \rec_i- \mathbf{X_i} ||_1 + (1-{\text{SSIM}}(\rec_i,\mathbf{X_i})) \right ),
\end{equation}
where SSIM is the structural similarity index~\cite{ssim2004}. Here $N$ represents the number of training images. However, solving~(\ref{eq:ltask}) may make the trajectory infeasible while learning the trajectory for the MRI task.  The gradient constraints are enforced by an additional constraint as proposed in~\cite{pilot}
\begin{equation}
	\label{eq:lconst}
	L_{\text{const}} = \lambda_v \sum_{i = 1}^{m-1} \max(0,|v_i|-v_{\max}) + \lambda_a \sum_{i = 1}^{m-2}  \max(0,|a_i|-a_{\max}),	
\end{equation}
with $v_i$ and $a_i$ as velocity and acceleration at $i^{th}$ point, calculated using first and second derivatives of the trajectory as $\dot{\mathbf{k}}$ and $\ddot{\mathbf{k}}$, respectively.

The jointly trained network $\Dk$ using the total loss function results in optimized trajectory $\mathbf k$ and network parameters $\theta$ for a particular decimation rate (DR). We can acquire undersampled measurements using this optimized trajectory $\mathbf k$ and later reconstruct the fully sampled image $\rec$ using this same network $\Dk$ as shown in the inference pipeline in Fig.~\ref{fig:model}(b).

\section{Experiments and Results}

We trained and tested our models on a subset of parallel imaging fastMRI knee dataset~\cite{kspace2018fastMRI} consisting of 100 volumes for training, 50 volumes for validation, and 94 volumes for testing. We performed network training on complex valued data at actual resolution without cropping. Images were cropped to center $320\times 320$ only for display purpose.  For simulating the single-shot data, we assumed, echo time~(TE) =100~ms, T2 for bone marrow fat=80~ms, and sampling duration= 1~$\mu s$.  During loss function optimization, we set the value of gradient constraints as $G_{\max} = 40$mT/m and $S_{\max} = 200$mT/m/ms.

\subsection{Improved Input to the Network}
\begin{figure}
	\centering	
	\includegraphics[width=0.9\linewidth]{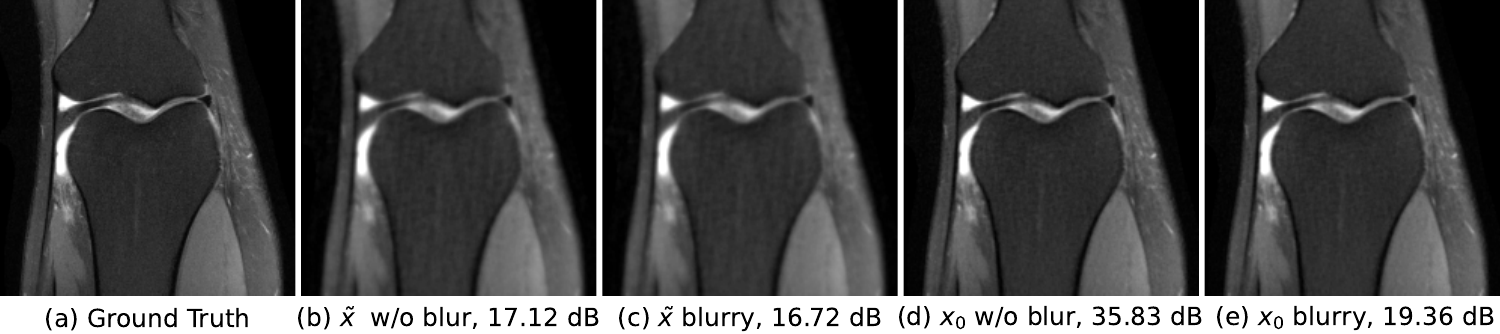}	
	\caption{This figure contrasts analytic and iterative reconstructions used as network inputs. A fully sampled test image shown in (a) when undersampled at 8x and reconstructed using  adjoint NUFFT with DCF results in (b). (c) and (d) depict iterative reconstructions via the SENSE algorithm, without and with blur in measurements, respectively. The blurred measurements lower the PSNR from 33.83 dB to 19.36 dB, highlighting the challenge of single-shot reconstruction. }
	\label{fig:sense}	
\end{figure}

Figure~\ref{fig:sense} highlights the impact of performing SENSE reconstruction using~\eqref{eq:sensesol} with and without~(w/o) considering the T2-Blur in the measurements. As we can see with a lower peak signal to noise ratio~(PSNR) value of 19.36~dB in Fig.~\ref{fig:sense}(d), the single-shot reconstruction problem becomes significantly challenging due to the presence of T2-blur. However, we note that the initial solution $\mathbf x_0$~(Fig.~\ref{fig:sense}(d)) has reduced artifacts compared to $\als$~(Fig.~\ref{fig:sense}(c)) and, therefore acts as a good initial input to the neural network. 

\subsection{Improved Reconstruction by the Network}
\begin{table} \centering
	\caption{Average PSNR (dB) and SSIM values at 8x and 16x acceleration factors shown as mean $\pm$ standard deviation for 94 test subjects. Typically, higher average PSNR and SSIM values signify enhanced reconstruction quality. When compared to the classical CSTV reconstruction and the PILOT algorithm, our LSST approach yields superior PSNR and SSIM values for single-shot acquisition. }
	\label{tab:8x16x}
	\begin{tabular}{|l|cl|ll|} \toprule
		& \multicolumn{2}{c|}{PSNR}                                       & \multicolumn{2}{c|}{SSIM $\times$ 100}                         \\ \midrule
		& 8x                                   & \multicolumn{1}{c|}{16x} & \multicolumn{1}{c}{8x} & \multicolumn{1}{c|}{16x} \\ \midrule
		CSTV    & $33.04 \pm 1.06$    & $30.91  \pm 1.23$    & $86.40 \pm 2.32$       & $82.46 \pm 2.76$        \\ \midrule	
		%No traj opt  & $36.81 \pm 1.48$    & $35.98  \pm 1.34$    & $92.67 \pm 2.67$       & $90.70 \pm 3.42$        \\ \midrule
		PILOT & $36.49 \pm 1.58$    & $ 35.41 \pm 1.44$    & $91.88 \pm 3.33$       & $90.27 \pm 3.78$        \\ \midrule
		Proposed  & $\mathbf{ 37.94 \pm 1.89}$    & $\mathbf {36.17  \pm 1.48}$    & $\mathbf{93.92 \pm 2.69}$       & $\mathbf{91.82 \pm 3.37}$    \\   \bottomrule
	\end{tabular}
\end{table}

\begin{table} \centering
	\caption{Average Likert scores obtained for 20 volumes on five-point scale with scores as 1 to 5 include 1--non-diagnostic quality, 2--poor diagnostic quality, 3--fair diagnostic quality, 	4--good diagnostic quality, 5--excellent diagnostic quality. }
	\label{tab:likert}
	\begin{tabular}{|lccccc|} \toprule
		& SNR & Artifacts & Resolution & Contrast & Overall \\ \midrule
		Avg.~of  20 volumes & 5   & 5         & 4          & 5        & 5      \\ \bottomrule
	\end{tabular}
\end{table}

We performed initial experiment using traditional compressed sensing algorithm~\cite{cstv} (1D undersampling) with a total variation regularization~(CSTV) without accounting for the T2-Blur at 8x accelerated data acquired using uniform Cartesian sampling mask having 18 lines in the calibration region.
In the second experiment, to ensure fair comparison, we extended the existing PILOT algorithm~\cite{pilot} to the single-shot settings accounting for T2-Blur in the measurements. 
In the third experiment, we implemented our proposed end-to-end training pipeline LSST as shown in Fig.~\ref{fig:model}(a).
% We simultaneously optimized trajectory and network parameters in single-shot acquisition settings where input to the network was $\mathbf x_0$ while accounting for DCF during regridding. 

Table~\ref{tab:8x16x} summarizes the PSNR and SSIM metrics on the test dataset at 8x and 16x acceleration factors for the three different methods previously discussed. Table~\ref{tab:likert} provides the quantitative evaluation results from an experienced radiologist on a five-point Likert scale. The evaluation metrics include SNR, Artifacts, Resolution, Contrast, and Overall Quality. 
It was observed that sharpness of ACL fibres was improved. The clinical implication for this is improved detection of partial ACL tears involving some of the fibres of the ACL.  The proposed method scored highest primarily due to improvements in the ACL fibres being sharper while the meniscus region was found to be uniformly sharp across the different methods.

\begin{figure}
	\centering	
	\includegraphics[width=.9\linewidth]{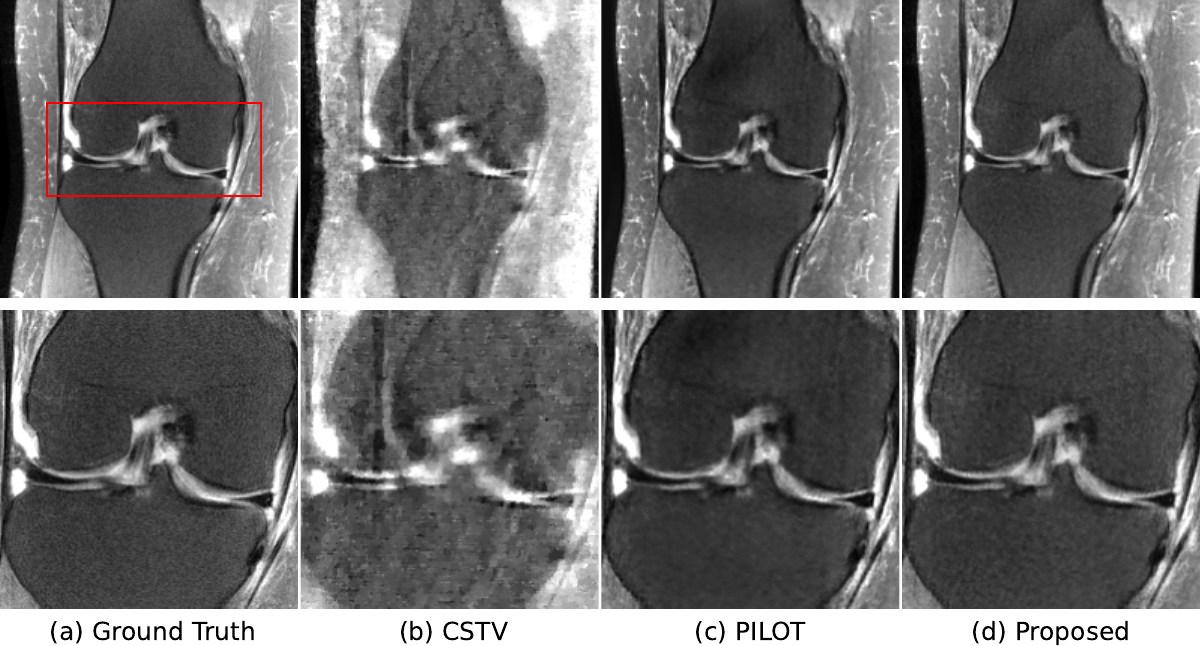}
	\caption{A comparison of experimental results at 8x acceleration on a test slice. The CSTV output~(b) exhibits noticeable artifacts due to high 8x acceleration and the use of a 1D Cartesian sampling mask. The zoomed area shows that ACL fibers are sharper than PILOT (c) whereas meniscus is sharp everywhere in the proposed method (d). }
	\label{fig:rec8x}
\end{figure}

Figure~\ref{fig:rec8x} visually compares the reconstruction quality of CSTV, PILOT and LSST at 8x acceleration on a slice of a test subject. The zoomed regions indicate the improvement in the reconstruction quality. Appendix shows additional results on trajectory evaluation at 8x and 16x acceleration factors. Figure~7 in the appendix shows benefits of individual SENSE and DL components of proposed pipeline.

\section{Conclusions and Discussions}
\label{Sec:discussion}

This study introduces a method for the joint optimization of the k-space trajectory and reconstruction network parameters for single-shot acquisition that accounts for T2-Blur while adhering to MR system constraints.  The learned trajectory meets gradient strength and slew-rate constraints and ensures practical trajectory estimation. Initial results and radiologist Likert scores affirm the method's utility. %Future plans include real data acquisition and reconstruction using the learned trajectory. The study shows that at high acceleration factors, the proposed framework with learned SSFSE acquisition improves reconstruction quality compared to 1D Cartesian under-sampling based FSE acquisition.

\section*{Acknowledgement}
We thank Shriram KS for his help and support with the manuscript.

\bibliographystyle{IEEEtran}
\bibliography{paperbib}

\section{Additional Results}

\begin{figure}
	\centering	
	\includegraphics[width=.95\linewidth]{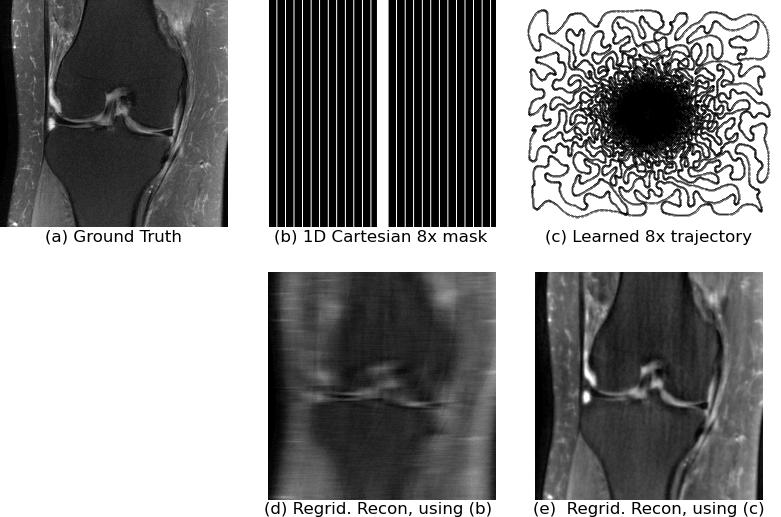}
	\caption{This figure shows the benefit of optimizing the trajectory rather than using traditional 1D Cartesian sampling to acquire the data. The ground truth image (a) was  undersampled using mask in (b) and trajectory in (c) that resulted in undersampled k-space data which was reconstructed using regridding reconstruction (regrid. recon.)~$\als$ to result in (d) and (e),respectively. Here, trajectory optimization also incorporated the T2-blur whereas Cartesian sampling experiment did not have additional blur.  We note that (d) and (e) are not the final reconstruction but zero-filled reconstructions~($\als$) as shown in Fig~1(a).  Since, (e) has many details as compared to (d), it can act as an improved input to the SENSE reconstruction block and subsequent neural network block.   }
	\label{fig:app8x}
\end{figure}

\begin{figure}
	\centering	
	\includegraphics[width=.99\linewidth]{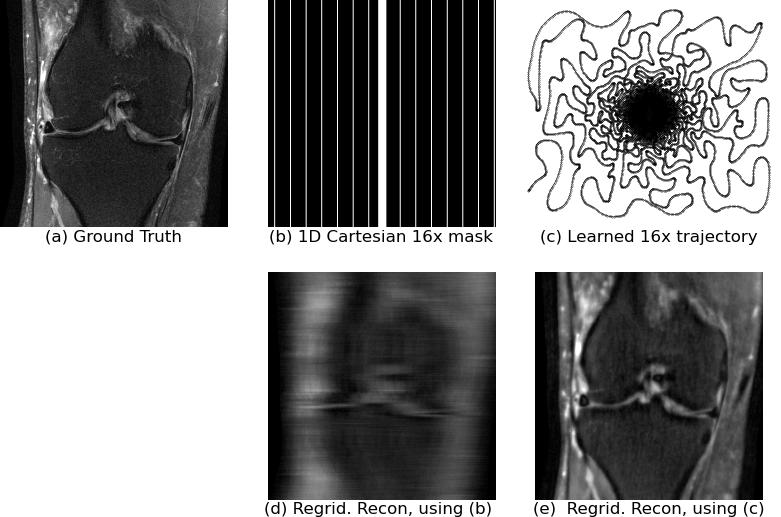}
	\caption{Same experiment as in Fig.~\ref{fig:app8x} but at 16x acceleration factor. It can be noted from (e) that even at the 16x acceleration factor, some details are visible in the initial NUFFT based reconstruction. Here again we note that (d) and (e) are not final reconstructions but the regridding reconstructions $\als$ that goes as input to the SENSE block and subsequently to neural network block as shown in Fig~1(a).}
	\label{fig:app16x}
\end{figure}

\begin{figure}
	\centering	
	\includegraphics[width=.99\linewidth]{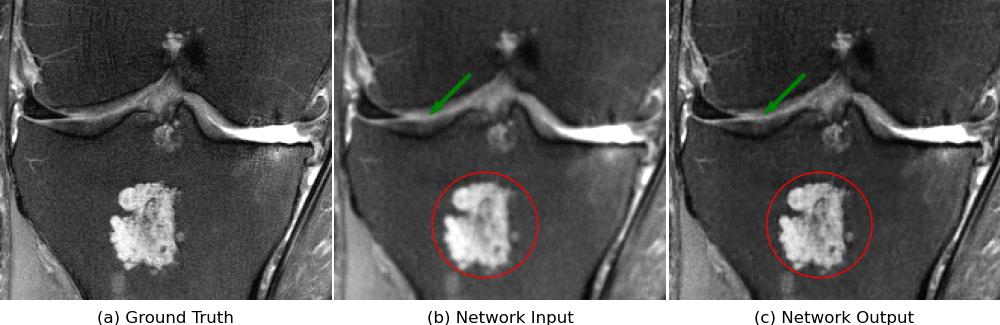}
	\caption{(a) is the ground truth fully sampled image. (b) is the output of SENSE algorithm on 8x undersampled and blurred measurements. The SENSE algorithm reduces the aliasing artifacts but does not reduce blur since the blur operator is unknown. This SENSE output acts as input to the direct-inversion network. (c) is the output of proposed network where blur is  reduced significantly. }
	\label{fig:appNetInOut}	
\end{figure}

\begin{figure}
	\centering	
	\includegraphics[width=.99\linewidth]{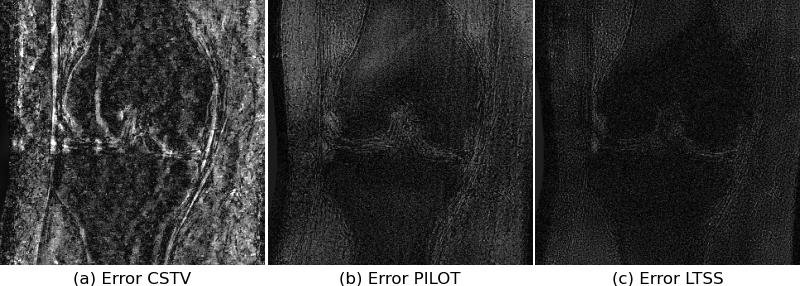}
	\caption{Error images between the ground truth image and reconstruction by different algorithms compared in Fig.4. The error maps are display range is adjusted for visualization purpose. }
	\label{fig:appError}	
\end{figure}

\end{document}